\documentclass[11pt,aps,pra]{revtex4}
\usepackage{graphicx}

\begin{document}

\title{Note on MacLennan-Zubarev Ensembles and QuasiStatic Processes
}
{

\author{Shuichi Tasaki$^a$ \\ Taku Matsui$^b$}

\address{$^a$Department of Applied Physics, Waseda University,\\
$^b$Graduate School of Mathematics,
Kyushu University}

{
\begin{abstract}\baselineskip=12pt
Relation between natural nonequilibrium steady states of Ruelle and MacLennan-Zubarev ensembles
is discussed for a system consisting of $M$ infinitely extended systems coupled with 
a small system. And a characterization of quasistatic processes is investigated
for a small system coupled with a single reservoir.
\end{abstract}
}
\maketitle

{\baselineskip=13pt
\section{Introduction}

The understanding of irreversible phenomena including nonequilibrium steady states is a
longstanding problem of statistical mechanics. 
Recent progress in the research of mesoscopic systems brings a new aspect 
into this problem. 
In these systems, coherence (a quantum dynamical aspect) may be observed
in a dissipative transport (an irreversible phenomenon) and
the two aspects should be discussed simultaneously. 
Usually, a mesoscopic system couples with much larger environments
and the interaction is not weak. As a result, the system cannot be 
clearly distinguished from the environments. 
Therefore, it is natural to deal with a mesoscopic system plus its 
environments as an infinitely extended system.

Statistical mechanics of infinitely extended systems has been developed
so far~\cite{Cornfeld,RuelleEq,Bratteli} and, recently, their 
nonequilibrium properties are studied intensively.
Those include analytical studies on nonequilibrium steady 
states of harmonic 
crystals~\cite{SpohnLeb1,Bafaluy},  a one-dimensional gas~\cite{Farmer}, 
unharmonic chains~\cite{Eckmann}, an 
isotropic XY-chain~\cite{HoAraki}, 
systems with asymptotic abelianness~\cite{Ruelle1}, 
a one-dimensional 
quantum conductor~\cite{ST},
an interacting fermion-spin system~\cite{JaksicPillet2},
fermionic junction systems~\cite{Frolich},
a quasi-spin model of superconductors~\cite{Verbeure} 
and a bosonic junction system
with Bose-Einstein condensates~\cite{STBose}.
Entropy production has been rigorously studied as well (see
Refs.~\cite{Ruelle1,JaksicPillet2,Ojima1,Ojima2,JaksicPillet1,RuelleEnPro}, 
and the references therein).
See also reviews~\cite{JaksicPillet3,STMatsui}.

For C$^*$-dynamical systems with $L^1$-asymptotic abelian properties, Ruelle\cite{Ruelle1} showed the
existence and stability of nonequlibrium steady states (NESS) which are naturally obtained 
from local equilibrium ensembles. We previously announced\cite{STMatsui} that, 
under stronger conditions, 
these natural NESS can be regarded as nonequilibrium ensembles proposed by MacLennan\cite{MacLennan}
and Zubarev\cite{Zubarev}. In this article, we give the proof of this statement.
On the other hand, one of us (ST) investigated\cite{STwaseda} a thermodynamic behavior 
of a driven small system coupled with an infinitely extended reservoir.
Another purpose of this article is
to give the detail proof and generalization of this observation. 
The rest part is arranged as follows: In the next section, we summarize the
setting. In Sec.~\ref{Sec:3}, the relation between natural NESS and Maclennan-Zubarev ensembles
is discussed. Sec.~\ref{Sec:4} is devoted to the discussions of the thermodynamic bahavior
of a small system coupled with an infinitely extended reservoir. Some remarks are given
in Sec.~\ref{Sec:5}.

\section{C$^*$-dynamical systems}\label{Sec:2}
\subsection{Field algebra}
The system ${\cal S}$ in question is described by a collection ${\cal F}$ of all finite
observables called a C$^*$-algebra, which is a complete linear space with 
a norm $\Vert \cdot \Vert$, where a procduct $AB$ and antilinear involution
*$:A\to A^*$ $(\forall A,B \in {\cal F}$) are defined and whose norm 
satisfies $\Vert AB \Vert \le \Vert A\Vert \Vert B\Vert$
and the C$^*$-property: $\Vert A^* A\Vert =\Vert A\Vert^2$.
We consider the case where an autonomous time evolution is defined which is 
described by a strongly continuous one-parameter 
group of *-automorphisms $\tau_t$ ($t\in {\bf R}$). Namely, $\tau_t$
is a linear map satisfying $\tau_t(AB)=\tau_t(A)\tau_t(B)$, $\tau_t(A^*)
=\tau_t(A)^*$, $\tau_0=I$ ($I$: the identity map), 
$\tau_t\tau_s=\tau_{t+s}$ and $\lim_{t\to 0}\Vert 
\tau_t(A)-A\Vert= 0$ ($^\forall A\in {\cal F}$).
Then, according to the theory of semigroups, there exists a 
densely defined generator $\delta$ of $\tau_t$:
$$
\lim_{t\to 0}\left\Vert \delta(A)-{1\over t}\{\tau_t(A)-A\}
\right\Vert=0 \ , \quad (\forall A\in D(\delta))
$$
where $D(\delta)$ is the domain of $\delta$.

We further assume that two more *-automorphisms are defined:
\smallskip

\begin{itemize}
\item [(i)] a strongly continuous $L$-parameter group of *-automorphisms $\alpha_{\vec \varphi}$ 
(${\vec \varphi}\in {\bf R}^L$) satisfying
$
\alpha_{{\vec \varphi}_1} \alpha_{{\vec \varphi}_2} = \alpha_{{\vec \varphi}_1+{\vec \varphi}_2}
\ , 
$
which represents the gauge transformation. 
\item [(ii)] an involutive *-automorphism $\Theta$ which is represented as 
$\Theta
=\alpha_{{\vec \varphi}_0}$ 
with some
${\vec \varphi}_0 \in {\bf R}^L$.
\end{itemize}
\smallskip

The automorphisms $\tau_t$, $\alpha_{\vec \varphi}$ and $\Theta$ are assumed to 
commute with each other:
$$
\Theta \tau_t = \tau_t \Theta, \quad \Theta \alpha_{\vec \varphi}=\alpha_{\vec \varphi} \Theta,
\quad \tau_t \alpha_{\vec \varphi} = \alpha_{\vec \varphi} \tau_t
\quad
( ^\forall t\in{\bf R} \ , ^\forall {\vec \varphi}\in{\bf R}^L) \ .
$$ 
A subalgebra ${\cal A} \subset {\cal F}$ consisting of invariant elements under the 
gauge transformations $\alpha_{\vec \varphi}$ (${\vec \varphi} \in {\bf R}^L$) is called the 
observable algebra, which describes observable physical quantities.
The *-automorphism $\Theta$ defines the even and odd subalgebras, respectively, 
${\cal F}_+$ and ${\cal F}_-$:
$$
{\cal F}_{\pm} = \{ A \in {\cal F} ; \Theta(A)=\pm A\} \ .
$$
When the system involves fermions, even and odd subalgebras correspond
to dynamical variables which are sums of products of, respectively, even and odd numbers of
fermion creation and/or annihilation operators.
Note that, since $\alpha_{s{\vec e}_\lambda}$ ($s\in {\bf R}$, ${\vec e}_\lambda$: 
a unit vector whose $\lambda$th element is unity) defines a strongly continuous group 
of *-automorphisms, it has a densely defined generator which is denoted as $g_\lambda$. 
The C$^*$-algebra ${\cal F}$ with these *-automorphisms is called a field algebra 
\cite{Bratteli,ArakiKishimoto}.

\subsection{Decomposition of the system and initial local equilibrium states}

We consider the situation where the system ${\cal S}$ can be decomposed into $M$ independent
infinitely extended subsystems ${\cal R}_j$ ($j=1,\cdots M$), which play a role of reservoirs, 
and a finite-degree-of-freedom subsystem ${\cal S}_0$ interacting with  all the others. 
More precisely, the algebra ${\cal F}$ is a tensor product of $M$
infinite dimensional subalgebras ${\cal F}_{j}$ 
($j=1,\cdots M$), corresponding to
${\cal R}_j$, and a finite dimensional subalgebra ${\cal F}_S$, 
corresponding to ${\cal S}_0$:
\begin{equation}
{\cal F} = {\cal F}_S \otimes {\cal F}_{1} \otimes \cdots \otimes {\cal F}_{M} \ ,
\label{tensor}
\end{equation}
such that the following conditions are satisfied: 

\begin{itemize}

\item [(S1)] \ There exists a gauge-invariant time evolution group $\tau^{V}_t$ 
($t\in {\bf R}$)
which is a perturbation of $\tau_t$ by a selfadjoint element $-V\in {\cal A}\cap D(\delta)$ and 
which is a  product of strongly continuous groups ${\tilde \tau}_t^{(j)}$ 
($j=1,\cdots M$)  
independently acting on subalgebras ${\cal F}_{j}$ ($j=1,\cdots M$)
\begin{equation}
\tau^V_t = {\tilde \tau}_t^{(1)} \cdots {\tilde \tau}_t^{(M)} \ .
\label{EvoDecomp}
\end{equation}
Namely, ${\tilde \tau}_t^{(j)}$'s commute with each other and leave the elements of other 
subalgebras invariant:\footnote{Throughout this article, the subalgebra
${\bf 1}_S\otimes {\bf 1}_1\otimes \cdots\otimes
{\cal F}_{k}\otimes \cdots \otimes {\bf 1}_M$, where ${\bf 1}_S$ and ${\bf 1}_{j}$ are
unities, respectively, of ${\cal F}_S$ and ${\cal F}_{j}$, is abbrebiated as
${\cal F}_{j}$. Similarly, 
${\cal F}_S\otimes {\bf 1}_1\otimes \cdots \otimes {\bf 1}_M$ as 
${\cal F}_S$.}
\begin{eqnarray}
{\tilde \tau}_t^{(j)}(A) &=& A \qquad \qquad  (^\forall A\in {\cal F}_{k}, \   k\not=j, \  \ k,j=1,\cdots M)
\label{Decomp1} \\
{\tilde \tau}_t^{(j)} {\tilde \tau}_s^{(k)} &=& {\tilde \tau}_s^{(k)} {\tilde \tau}_t^{(j)}
\quad \ (t,s \in {\bf R}, \ k\not= j, \  \ k,j=1,\cdots M) \ . \label{Decomp2} 
\end{eqnarray}

\item [(S2)]  The gauge *-automorphism $\alpha_{\vec \varphi}$ is a product of strongly 
continuous 
groups ${\tilde \alpha}_{\vec \varphi}^{(j)}$ 
\break
($j=1,\cdots~M$)  
and ${\tilde \alpha}_{\vec \varphi}^S$: 
\begin{equation}
\alpha_{\vec \varphi} = {\tilde \alpha}_{{\vec \varphi}}^{S} {\tilde \alpha}_{\vec
\varphi}^{(1)} \cdots {\tilde \alpha}_{\vec
\varphi}^{(M)}
\ ,
\label{GauDecomp}
\end{equation}
where
${\tilde \alpha}_{\vec \varphi}^{(j)}$ 
and ${\tilde \alpha}_{\vec \varphi}^S$
independently act on subalgebras ${\cal F}_{j}$ ($j=1,\cdots~M$)
and ${\cal F}_S$
\begin{eqnarray}
{\tilde \alpha}_{\vec \varphi}^{(j)}(A) &=& A \qquad \qquad (
^\forall A\in 
{\cal F}_{k},
\   k\not=j, \  k,j=S,1,\cdots M)
\label{Decomp3} \\
{\tilde \alpha}_{{\vec \varphi}_1}^{(j)} {\tilde \alpha}_{{\vec \varphi}_2}^{(k)} &=& {\tilde
\alpha}_{{\vec
\varphi}_2}^{(k)} {\tilde \alpha}_{{\vec \varphi}_1}^{(j)}
\quad \  ({\vec \varphi}_1,{\vec \varphi}_2 \in {\bf R}^L, \ k\not= j, \  k,j=S,1,\cdots M) 
\label{Decomp4} 
\end{eqnarray}

\end{itemize}
\smallskip

\noindent
And the groups ${\tilde \tau}_t^{(j)}$ and ${\tilde \alpha}_{{\vec \varphi}}^{(k)}$ 
commute with each other:
$$
{\tilde \tau}_t^{(j)} {\tilde \alpha}_{{\vec \varphi}}^{(k)} = {\tilde
\alpha}_{{\vec \varphi}}^{(k)} {\tilde \tau}_t^{(j)} 
\quad (j,k=1,\cdots, M; \ t\in{\bf R},\ {\vec \varphi}\in{\bf R}^L) \ .
$$

States are introduced by listing expectation values.
Namely, each state is identified with a linear map $\omega$ from $A\in {\cal F}$
to an expectation value $\omega(A)$.
The positivity condition $\omega(A^*A)\ge 0$
and normalization condition $\omega({\bf 1})=1$ (with ${\bf 1}\in {\cal F}$ 
the unity) are required. 
As in the previous
works\cite{SpohnLeb1,HoAraki,ST,JaksicPillet2,Ojima1,Ojima2,JaksicPillet1}, we are
interested in  the evolution of initial states where $M$ infinitely extended subsystems 
(reservoirs) are in equilibrium with different temperatures and different chemical potentials 
and the finite subsystem is in an arbitrary non-singular state. As discussed in
\cite{Ojima2,JaksicPillet1}, such states are specified as a KMS state:
\begin{itemize}

\item[(S3)] Let $\sigma_x^{\omega}$ ($x\in {\bf R}$) be a strongly continuous group
defined by
\begin{equation}
\sigma_x^{\omega}(A) = \prod_{j=1}^M {\tilde\tau}_{-\beta_j x}^{(j)} 
{\tilde \alpha}_{\beta_j{\vec
\mu}_jx}^{(j)}\left(e^{i D_S x}A e^{-iD_S x}\right) \ , \quad (A\in {\cal F}) 
\label{KMS}
\end{equation}
where $\beta_j$ and ${\vec \mu}_j = (\mu_j^{(1)},\cdots \mu_j^{(L)})$ are, respectively, 
the inverse temperature and the set of chemical potentials of the $j$th reservoir. 
The operator
$D_S$ $(\in {\cal F}_S \cap {\cal A}$) is selfadjoint.
Then an initial state $\omega$, which
we are interested in, is a 
KMS state with temperature $-1$ with respect to $\sigma_x^{\omega}$.
Namely, $\omega$ is a state such that, for any pair $A,B\in {\cal F}$, there exists a
function $F_{A,B}(x)$ of $x$-analytic in the stripe $\{ x\in {\bf C}; 0> {\rm Im} x >-1\}$
and satisfies the KMS boundary condition:
\begin{equation}
F_{A,B}(x) = \omega(A\sigma_x^{\omega}(B)) \qquad F_{A,B}(x-i) =
\omega(\sigma_x^{\omega}(B) A) \quad (x\in {\bf R}) \ .
\end{equation}

\end{itemize}

\medskip

Hereafter, we assume that the domains of the generators ${\tilde \delta}_j$ and ${\tilde
g}_\lambda^{(j)}$, respectively, of 
${\tilde \tau}_t^{(j)}$ and
${\tilde \alpha}^{(j)}_{s{\vec e}_\lambda}$ ($t,s \in {\bf R}$) contain the domain
$D(\delta)$ of the generator $\delta$ of $\tau_t$:
\smallskip

\begin{itemize}
\item[(S4)] $D(\delta) \subset D({\tilde \delta}_j) \quad D(\delta) \subset D({\tilde
g}_\lambda^{(j)}) \qquad( ^\forall j=0,1,\cdots M$, $^\forall \lambda = 1,\cdots L$).
\end{itemize}
\medskip

\noindent
Then, the condition (S1) implies that the domain of the generator ${\delta}^V$ of $\tau^V_t$
is identical with $D(\delta)$: $D({\delta}^V)=D(\delta)$ and
\begin{eqnarray}
\delta(A) &=& {\delta}^V(A) + i [V,A] \quad  \left(A\in D(\delta)\right) \\
{\delta}^V(A)&=& \sum_{j=1}^N {\tilde \delta}_j(A) \qquad \quad \ \left(A\in
D(\delta)\right)
\end{eqnarray}
Moreover, the domain of the generator ${\hat \delta}_\omega$ of $\sigma_x^{\omega}$
also includes $D(\delta)$ and
\begin{equation}
{\hat \delta}_\omega(A) = - \sum_{j=1}^N \left\{ \beta_j \left({\tilde \delta}_j(A) -
\mu_\lambda^{(j)} {\tilde g}_\lambda^{(j)}(A)\right)\right\} + i [D_S, A] \ . \quad (A\in
D(\delta)) \label{SigmaGener}
\end{equation}
Note that a decomposition without the finite subsystem is possible
as well.

\subsection{$L^1$-asymptotic abelian property}
In thermodynamics, environments are assumed to stay in equilibrium 
under arbitrary processes and their details are considered to be 
unimportant. 
Hence, thermodynamic environments would be well-modelled by 
systems with appropriate ergodicity. As one of such an example,
we consider systems satisfying the $L^1$-asymptotic abelian property.

The time evolution $\tau_t$ is said to satisfy the 
$L^1({\cal G})$-asymptotic abelian property if 
there exists a norm dense *-subalgebra ${\cal G}$ such that
\begin{eqnarray}
\int_{-\infty}^{+\infty} dt \Vert [A,\tau_t(B)]\Vert
\equiv
\int_{-\infty}^{+\infty} dt \Vert A\tau_t(B)-\tau_t(B)A \Vert &<& +\infty \quad (A\in {\cal G}, \ 
B\in {\cal G}\cap {\cal F}_+)  \cr
\int_{-\infty}^{+\infty} dt \Vert [A,\tau_t(B)]_+\Vert 
\equiv
\int_{-\infty}^{+\infty} dt \Vert A\tau_t(B)+\tau_t(B)A \Vert
&<& +\infty \quad (A, B\in
{\cal G}\cap {\cal F}_-) \nonumber
\end{eqnarray}
where ${\cal F}_\pm$ are even/odd subalgebras.
This property implies rapid decay of correlations and is satisfied by
free fermions in ${\bf R}^d$ ($d\ge 1$) (Example 5.4.9 of Ref.~\cite{Bratteli}).
Note that, if a system admits bound states, it does not satisfy the 
$L^1$-asymptotic abelian condition as there exist bounded constants 
of motion, i.e., observables $C$ satisfying $\tau_t(C)=C$.

\section{NESS
and MacLennan-Zubarev ensembles}\label{Sec:3} 

The state $\omega_t$ at time $t$ starting from the initial state $\omega$ is given by
\begin{equation}
\omega_t(A) = \omega\left(\tau_t(A)\right) \ , \quad (^\forall A\in{\cal F}) \label{StatEvo}
\end{equation}
and, under the setting (S1)-(S4), nonequlibrium steady states are expected to
be obtained as its weak limits for $t\to \pm \infty$.
As shown by Ruelle\cite{Ruelle1}, it is indeed the case. Namely, under the setting (S1)-(S4),
if the time evolution $\tau_t$ is $L^1({\cal G})$-asymptotic abelian and the perturbation $V$
is an element of $\cal G$, the limits
$$
\lim_{t\to\pm \infty}\omega\left(\tau_t(A)\right)=\omega(\gamma_\pm(A))\equiv 
\omega_\pm(A)
$$ 
exist for all $A\in {\cal F}$ and define nonequilibrium steady states, where $\gamma_\pm$
are M\o ller morphisms defined by
$$
\lim_{t\to \pm\infty}\left\Vert {\tau^V_t}^{-1}\tau_t(A)-\gamma_\pm(A)
\right\Vert=0 \quad (^\forall A\in {\cal F}) \ .
$$
Also Ruelle showed the independence of the limits on the way of separating the whole system 
into reservoirs and a small system\cite{Ruelle1}. See also Refs.~\cite{JaksicPillet2,Frolich}.
In this section, as announced in Ref.~\cite{STMatsui}, 
under a stronger condition, these steady states $\omega_\pm$ are shown to be
inerpreted as MacLennan-Zubarev ensembles\cite{MacLennan,Zubarev}.
Indeed, one has:
\medskip

\noindent{\it Proposition 1: KMS characterization of evolving states}
\begin{quote}
Under the setting (S1)-(S4), 
the state $\omega_t$ at time $t$ is a KMS state at temperature $-1$ with 
respect to the strongly continuous group of *-automorphisms 
\begin{equation}
\sigma_x^{\omega_t} \equiv \gamma_t^{-1} \sigma_x^{\omega} \gamma_t \ ,
\end{equation}
where $\gamma_t = {\tau_t^V}^{-1}\tau_t$, and its generator is given by 
\begin{equation}
{\hat \delta}_\omega^{(t)}(A) = {\hat \delta}_\omega(A) + i \int_{-t}^0 ds 
\left[ \tau_s\left({\hat \delta}_\omega(V)\right),A\right] \ ,
\quad 
( ^\forall A\in D({\hat \delta}_\omega^{(t)})=D({\hat \delta}_\omega)) \ .
\end{equation}
\end{quote}

\medskip

\noindent
{\it Proposition 2: KMS characterization of steady states}
\begin{quote}
Under the setting (S1)-(S4), if 
the time evolution *-automorphism $\tau_t$ is $L^1({\cal G})$-asymptotically abelian,
$V\in {\cal G}$
and the M\o ller morphisims $\gamma_\pm$ are invertible, the steady states 
$\omega_\pm$ are KMS states at temperature $-1$ with 
respect to the strongly continuous group of *-automorphisms 
\begin{equation}
\sigma_x^{\omega_\pm} \equiv \gamma_\pm^{-1} \sigma_x^{\omega} \gamma_\pm \ .
\end{equation}
Furthermore, if ${\hat \delta}_\omega(V) \in {\cal G}$, its generator 
${\hat \delta}_\omega^\pm$ satisfies
\begin{equation}
{\hat \delta}_\omega^\pm(A) = {\hat \delta}_\omega(A) + i \int_{\mp \infty}^0 ds 
\left[ \tau_s\left({\hat \delta}_\omega(V)\right),A\right] \ , 
\quad (^\forall A\in D({\hat \delta}_\omega)\cap {\cal G}) \ .
\label{SteadyKMS}
\end{equation}
\end{quote}

\medskip

\noindent
{\bf NB 3} \ For finite systems, the KMS state $\omega$ with respect to the *-automorphism
$\sigma_x^{\omega}$ corresponds to the density matrix
$$
\rho_\omega = {1\over Z}\exp\Bigl\{-\sum_{j=1}^N \beta_j \Bigl(H_j - \sum_{\lambda=1}^L
\mu_\lambda^{(j)}  N^{(\lambda)}_j \Bigr) \Bigr\} \ ,
$$
where $Z$ is the normalization constant, $\beta_j$, $H_j$, $\mu_\lambda^{(j)}$ and $N^{(\lambda)}_j$
are, respectively, the local temperature, local energy, local chemical potential and local
number operator of the $j$th reservoir. 
Then, as a result of the Liouville-von Neumann eqation, the density matrix at time $t$
is given by
\begin{eqnarray}
\tau_t^{-1}(\rho_\omega)&=& {1\over Z}\exp\Bigl\{-\sum_{j=1}^M \beta_j \Bigl( \tau_{-t}( H_j) 
- \sum_{\lambda=1}^L
\mu_\lambda^{(j)}  \tau_{-t}(N^{(\lambda)}_j) \Bigr) \Bigr\} \cr
&=&
{1\over Z}\exp\Bigl\{-\sum_{j=1}^M \beta_j \Bigl[H_j - \sum_{\lambda=1}^L
\mu_\lambda^{(j)}  N^{(\lambda)}_j - \int_{-t}^0 ds \tau_s(J_j^q)
\Bigr] \Bigr\} \ , \nonumber
\end{eqnarray}
where 
$
J_j^q \equiv {\displaystyle{d\over dt}}\tau_t\left(H_j\right) - \sum_{\lambda=1}^L
\mu_\lambda^{(j)}{\displaystyle{d\over dt}}\tau_t\bigl(N_j^{(\lambda)} \bigr)\Big|_{t=0}
= -i [H_j,V] + \sum_{\lambda=1}^L \mu_\lambda^{(j)} i[N_j^{(\lambda)},V] \ 
$
is non-systematic energy flow, or heat flow, to the $j$th reservoir
and we have used 
\begin{eqnarray}
\tau_{-t}(H_j) = H_j-\int_{-t}^0 ds 
{d\over ds}\tau_{s}(H_j) \ , \quad
\tau_{-t}(N^{(\lambda)}_j) = N^{(\lambda)}_j-\int_{-t}^0 ds 
{d\over ds}\tau_{s}(N^{(\lambda)}_j) \ . \nonumber 
\end{eqnarray}
On the other hand, as ${\tilde\delta}_j(V)=i[H_j,V]$ and ${\tilde
g}_\lambda^{(j)}=i[N_j^{(\lambda)},V]$, one has
$$
\sum_{j=1}^M \beta_j \tau_s(J_j^q)= -\tau_s\Bigl(\sum_{j=1}^M \beta_j 
\{{\tilde\delta}_j(V)-\sum_{\lambda=1}^L \mu_\lambda^{(j)} 
{\tilde g}_\lambda^{(j)}(V)\}\Bigr)
=-
\tau_s \left({\hat \delta}_\omega(V)\right)
$$
and, thus, 
$$
\tau_t^{-1}(\rho_\omega)
=
{1\over Z}\exp\Bigl\{-\sum_{j=1}^M \beta_j \Bigl[H_j - \sum_{\lambda=1}^L
\mu_\lambda^{(j)}  N^{(\lambda)}_j + \int_{-t}^0 ds 
\tau_s \left({\hat \delta}_\omega(V)\right)
\Bigr] \Bigr\} \ , 
$$
which is a KMS state generated by
$
\displaystyle
{\hat \delta}_\omega(A)
+i\int_{-t}^0 ds 
[\tau_s \left( {\hat \delta}_\omega (V)\right),
A] \ .
$
This observation is nothing but the finite dimensional version of Proposition 1. 

\medskip

\noindent
{\bf NB 4} \  For infinite systems, an interesting case is that where the right-hand side
of (\ref{SteadyKMS}) generates $\sigma_x^{\omega \pm}$. 
Then, if the integral 
\begin{equation}
{\widetilde V}_\pm \equiv \int_{\mp \infty}^0 ds \tau_s\left({\hat \delta}_\omega(V)\right)
\label{Integral}
\end{equation}
would converge, $\omega_\pm$ would be perturbed KMS states of the initial state $\omega$
by self-adjoint operators ${\widetilde V}_\pm$.
Moreover, NB 3 suggests that the corresponding density matrices would be
\begin{equation}
\rho_{\pm} = {1\over Z}\exp\Bigl\{-\sum_{j=1}^N \beta_j \Bigl[H_j - \sum_{\lambda=1}^L
\mu_\lambda^{(j)}  N^{(\lambda)}_j - \int_{\mp \infty}^0 ds \tau_s(J_j^q)
\Bigr] \Bigr\} \ . \label{MacLennanZubarevDM}
\end{equation}
Note that such ensembles for steady states were introduced by MacLennan\cite{MacLennan}
and Zubarev\cite{Zubarev}. 

However, if the steady states carry nonvanishing entropy production, the integral 
${\widetilde V}_\pm$ does not converge since the steady-state average of its integrand 
is nothing but the nonvanishing enropy production rate at the steady states:
$$
\lim_{s\to \pm \infty} \omega\Bigl(\tau_s\left({\hat \delta}_\omega(V)\right)\Bigr)
= \omega_\pm\Bigl(\sum_{j=1}^M\beta_jJ_j^q\Bigr)
\not =0 \ .
$$
This observation is consistent with the results by Jak\v si\' c and Pillet\cite{JaksicPillet1} who showed that,
if the steady-state entropy production is nonvanishing, the steady state is `singular' with
respect to the initial local equilibrium state.
Thus, the original proposal (\ref{MacLennanZubarevDM}) by MacLennan and Zubarev
cannot be justified. Rather, the KMS states with respect to $\sigma_x^{\omega \pm}$ 
generated by
(\ref{SteadyKMS}) should be regarded as a precise definition of the MacLennan-Zubarev 
ensembles.

\medskip

\noindent {\bf Proof of Proposition 1}

As easily seen from (S1) and (S4), the initial state $\omega$ is invariant under
$\tau^V_t$. And one has
$
\omega_t(A) = \omega\left(\tau_t(A)\right) = \omega\left(\gamma_t(A)\right) \quad (^\forall A\in {\cal F})\ .
$
Since $\gamma_t$ is a *-automorphism and $\omega$ is a KMS state with respect to 
$\sigma_x^\omega$ at temperature $-1$, $\omega_t$ is a KMS state with respect to 
$\gamma_t^{-1} \sigma_x^\omega
\gamma_t$ at temperature $-1$ (cf. Prop. 5.3.33 of Ref.~\cite{Bratteli}). 

Now we consider the generator.
In terms of the one-parameter family $Y_t$ of unitary elements 
defined as a norm convergent series:
\begin{equation}
Y_t = {\bf 1} + \sum_{n=1}^{+\infty} i^n \int_0^t dt_1 \int_0^{t_1} dt_2 \cdots \int_0^{t_{n-1}} dt_n
\tau_{-t_n}(V)\cdots \tau_{-t_2}(V) \tau_{-t_1}(V)
\ , \label{Cocycle}
\end{equation}
one has $\gamma_t(A)=Y_t A Y_t^*$.
Thus, $\gamma_t(A)$ ($ ^\forall A\in {\cal F}$) is differentiable with respect to $t$ and
$\displaystyle
{d\over dt}\gamma_t(A) = i \gamma_t\left([\tau_{-t}(V),A]\right) \ .
$
This leads to
$$
{d \{\gamma_t^{-1} \sigma_x^\omega \gamma_t(A)\} \over dt} 
=
i [\gamma_t^{-1} \sigma_x^\omega \gamma_t \tau_{-t}(V)-\tau_{-t}(V),\gamma_t^{-1} 
\sigma_x^\omega \gamma_t(A)] 
= i \left[\tau_{-t}\left(\sigma_x^\omega(V)-V\right),\gamma_t^{-1} 
\sigma_x^\omega \gamma_t(A)\right]
$$
or
\begin{equation}
\gamma_t^{-1} \sigma_x^\omega \gamma_t(A)
= \sigma_x^\omega(A)+i \int_0^t ds \left[\tau_{-s}\left(\sigma_x^\omega(V)-V\right),\gamma_s^{-1} 
\sigma_x^\omega \gamma_s(A)\right] \ ,
\end{equation}
where $\gamma_t \tau_{-t}=\tau_{-t}^V$, $\tau_{-t}^V\sigma_x^\omega=\sigma_x^\omega
\tau_{-t}^V$ and $\gamma_{t}^{-1} \tau_{-t}^V =\tau_{-t}$ have been used. Since 
$V\in~D(\delta)\subset~D({\hat\delta}_\omega)$, one has
the following relation for any $A\in D({\hat \delta}_\omega)$:

\begin{eqnarray}
&&\left.{d\over dx}\gamma_t^{-1} \sigma_x^\omega \gamma_t(A)\right|_{x=0}\nonumber \\
&&= \lim_{x\to 0} \left\{{\sigma_x^\omega(A) - A\over x}+i \int_0^t ds
\left[\tau_{-s}\left({\sigma_x^\omega(V)-V\over x}\right),\gamma_s^{-1} 
\sigma_x^\omega \gamma_s(A)\right]\right\} \nonumber \\
&&= {\hat \delta}_\omega(A)+i \int_{-t}^0 ds
\left[\tau_{s}\left({\hat \delta}_\omega(V)\right),A \right] \ . \label{GenSug}
\end{eqnarray}
This suggests that ${\widetilde \delta}_{\omega_t}$ defined by
$$
{\widetilde \delta}_{\omega_t}(A) \equiv {\hat \delta}_\omega(A)+i \int_{-t}^0 ds
\left[\tau_{s}\left({\hat \delta}_\omega(V)\right),A \right]
$$
is the generator ${\hat \delta}^{(t)}_\omega$ of $\gamma_t^{-1} \sigma_x^\omega \gamma_t$. 
As we see,
it is the case. First, we note that, since $\int_{-t}^0 ds
\tau_{s}\left({\hat \delta}_\omega(V)\right)$ is selfadjoint, ${\widetilde \delta}_{\omega_t}$
generates a strongly continuous group (cf. Theorem~4.1 of Ref.~\cite{Bratteli}
). Then as a
result of the theory of semigroups (Prop. 3.1.6 of Ref.~\cite{Bratteli}
), its resolvent
$
( \mu 1 + {\widetilde \delta}_{\omega_t})^{-1}
$
($\mu \in {\bf R}\backslash \{0\}$) is bounded, where $1$ stands for the identity operator
on ${\cal F}$. For any $B\in {\cal F}$, let $A=( \mu 1 + {\widetilde
\delta}_{\omega_t})^{-1}B$, then $A\in D({\hat \delta}_\omega)$ and (\ref{GenSug}) leads to
$$
(\mu 1 + {\hat \delta}^{(t)}_\omega) A = \mu A + \left.{d\over dx}\gamma_t^{-1} \sigma_x^\omega
\gamma_t(A)\right|_{x=0} =B \ .
$$
On the other hand, as a generator of strongly continuous group $\gamma_t^{-1} \sigma_x^\omega
\gamma_t$, the resolvent $(\mu 1 + {\hat \delta}^{(t)}_\omega)^{-1}$ of ${\hat 
\delta}^{(t)}_\omega$ is again bounded. Therefore, one finally has
$$
(\mu 1 + {\hat \delta}^{(t)}_\omega)^{-1} B = A =( \mu 1 + {\widetilde
\delta}_{\omega_t})^{-1}B
$$
for all $B\in {\cal F}$ and ${\hat \delta}^{(t)}_\omega={\widetilde
\delta}_{\omega_t}$. Moreover, as $D({\widetilde
\delta}_{\omega_t})=D({\hat \delta}_\omega)$, $D({\hat \delta}^{(t)}_\omega)=D({\hat
\delta}_\omega)$.

\medskip

\noindent
{\bf Proof of Proposition 2}

Remind that $\omega_{\pm}(A)=\omega \left( \gamma_{\pm}(A)\right)$ ($^\forall A\in {\cal F}$). 
When the M\o ller
morphisms $\gamma_{\pm}$ are invertible, they are *-automorphisms. Thus, since 
$\omega$ is a KMS
state with respect to $\sigma_x^\omega$ at temperature $-1$, $\omega_\pm$ is a KMS 
state with respect
to $\gamma_\pm^{-1} \sigma_x^\omega
\gamma_\pm$ at temperature $-1$ (cf. Prop. 5.3.33 of Ref.~\cite{Bratteli}). 

Now we show the second half. Since *-automorphisms are norm preserving 
(cf. e.g. Corollary 2.3.4 of Ref.~\cite{Bratteli}
), one has
\begin{eqnarray}
&&\Vert \gamma_{\pm n}^{-1} \sigma_x^\omega
\gamma_{\pm n}(A) - \gamma_\pm^{-1} \sigma_x^\omega \gamma_\pm(A)\Vert
\le \Vert \gamma_{\pm n}(A)-\gamma_\pm(A)\Vert + F(x,n) \ ,
\end{eqnarray}
where $\gamma_{\pm n}={\tau^V_{\pm n}}^{-1}\tau_{\pm n}$ and
$$
F(x,n)=\left\Vert \gamma_{\pm n}\left( \gamma_\pm^{-1} \sigma_x^\omega
\gamma_\pm(A)\right)- \gamma_{\pm}\left( \gamma_\pm^{-1} \sigma_x^\omega
\gamma_\pm(A)\right)\right\Vert \ .
$$
As the M\o ller morphism is a strong operator 
limit of $\gamma_{\pm n}$, 
\begin{eqnarray}
\lim_{n\to +\infty} \Vert \gamma_{\pm n}(A)-\gamma_\pm(A)\Vert &=&0 , 
\qquad (^\forall A\in {\cal F})
\\
\lim_{n\to +\infty}F(x,n)&=&0 \ . \qquad ({\rm for \ each \ } x )\label{UniLim}
\end{eqnarray}
Next, as a result of
\begin{eqnarray}
|F(x,n)-F(x',n)|&\le& \Vert \gamma_{\pm n} \gamma_\pm^{-1}\left\{ 
\sigma_x^\omega \gamma_\pm(A)
- \sigma_{x'}^\omega \gamma_\pm(A) \right\}\Vert \nonumber \\
&=& \Vert  \sigma_x^\omega \gamma_\pm(A)
- \sigma_{x'}^\omega \gamma_\pm(A) \Vert \ , \nonumber
\end{eqnarray}
the function $x \to F(x,n)$ is continuous uniformly with respect to $n$. Hence, the limit
(\ref{UniLim}) is uniform with respect to $x$ on any finite interval in ${\bf R}$. 
In short, 
\begin{eqnarray}
\lim_{n\to +\infty}\Vert \gamma_{\pm n}^{-1} \sigma_x^\omega
\gamma_{\pm n}(A) - \gamma_\pm^{-1} \sigma_x^\omega \gamma_\pm(A)\Vert =0 \ ,
\end{eqnarray}
where the convergence is uniform with respect to $x$ on any finite interval in ${\bf R}$.
Hence, as a result of Theorem 3.1.28 of Ref.~\cite{Bratteli}
, the generator ${\hat
\delta}_\omega^\pm$ of $\gamma_\pm^{-1} \sigma_x^\omega \gamma_\pm$ is the graph 
limit of the generators 
${\hat \delta}^{(\pm n)}_\omega$ of $\gamma_{\pm n}^{-1} \sigma_x^\omega \gamma_{\pm n}$.
Namely, if a sequence $\{ A_n \}_{n=1}^{+\infty}$ with $A_n \in 
D({\hat \delta}^{(\pm n)}_\omega)$
satisfies $A_n\to A$ and ${\hat \delta}^{(\pm n)}_\omega(A_n) \to B$ ($n\to +\infty$) , 
then one has $B={\hat \delta}_\omega^\pm(A)$.

For each $A\in D({\hat \delta}_\omega)\cap {\cal G}$, let $A_n =A$, then obviously $\lim_{n\to
+\infty} A_n=A$, $A_n \in D({\hat \delta}^{(\pm n)}_\omega)$ because of Proposition 1. And, 
as ${\hat
\delta}_\omega(V) \in {\cal G}$ is assumed, the $L^1({\cal G})$-asymptotic abelian property of $\tau_t$
implies
$$
\lim_{n\to +\infty} {\hat \delta}^{(\pm n)}_\omega(A_n) = {\hat \delta}_\omega(A)+i 
\int_{\mp \infty}^0 ds \left[\tau_{s}\left({\hat \delta}_\omega(V)\right),A \right] \ .
$$
Therefore, one has the desired result
$$
{\hat \delta}_\omega^\pm(A) =  {\hat \delta}_\omega(A)+i 
\int_{\mp \infty}^0 ds \left[\tau_{s}\left({\hat \delta}_\omega(V)\right),A \right] \ .
$$

\section{Quasistatic process and Clausius equality}\label{Sec:4}

In this section, we consider the case where ${\cal F}_S$ is the algebra of bounded
operators on a finite dimensional Hilbert space and the small system couples with a single 
reservoir, namely the case of $M=1$. Then, the total system is described by the tensor product
${\cal F}_S\otimes {\cal F}_{1}$ and the `decoupled' evolution $\tau^V_t$ acts only 
on the reservoir algebra: $\tau^V_t\equiv {\tilde \tau}_t^{(1)}$. 
Since the system is finite dimensional, the generator
of the gauge transformation ${\tilde\alpha}_{s{\vec e}_\lambda}^{(S)}$ is a commutator with 
a self-adjoint element $N_\lambda\otimes {\bf 1}_1 $
where $N_\lambda\in{\cal F}_S$ is the number operator of 
the $\lambda$th particles and ${\bf 1}_1$ is the unit of ${\cal F}_1$. 
The system-reservoir interaction is assumed to be proportional to a 
coupling constant $\kappa$: $\kappa V$ where $V\in D(\delta)$ is self-adjoint and gauge-invariant: 
$\alpha_{\vec \varphi}(V)=V$. Note that the gauge 
invariance of $V$ leads to ${\tilde g}_\lambda^{(1)}(V)+i[N_\lambda\otimes {\bf 1}_1,V]=0$
($\lambda=1,\cdots L$).

We are interested in the response of the whole system under a time-dependent 
perturbation:~$W(t)\otimes {\bf 1}_1$ where $W(t)\in {\cal F}_S$ is twice 
continuously differentiable in norm, $W(t)=W_0$ for $t\le 0$ and $[N_\lambda\otimes {\bf 1}_1,
W(t)\otimes {\bf 1}_1]=0$.
Initially, the whole system is prepared to be an equilibrium state $\omega$ of the inverse
temperature $\beta$ and the chemical potentials $\mu_\lambda$, namely, a
KMS state at $\beta$ with respect to ${\hat \sigma}_s\equiv \alpha_s^{(i)}\alpha_{-s {\vec \mu}}$
where $\alpha_t^{(i)}$ is defined by
$$
{d{\alpha}^{(i)}_t(A)\over dt} ={\alpha}^{(i)}_t\left(
{\tilde\delta}_1(A)+i[ W_0\otimes {\bf 1}_1+\kappa V,A]\right) \ ,
\quad {\alpha}^{(i)}_t(A)|_{t=0}=A \ ,
\quad (^\forall A\in D({\tilde \delta}_1) )
\ .
$$
The time evolution $\tau^W_t$ is given by the solution of
$$
{d\tau^W_t(A)\over dt} =\tau^W_t\left(
{\tilde\delta}_1(A)+i[ \kappa V+W(t)\otimes {\bf 1}_1,A]\right) \ ,
\quad \tau^W_t(A)|_{t=0}=A \ ,
\quad (A\in D({\tilde \delta}_1) )
\ ,
$$
and the state at time $t$ by $\omega_t(A)\equiv \omega\left(\tau^W_t(A)\right)$
($^\forall A\in {\cal F}$).
Now, we define the the system-energy increase $Z_T$ induced by the
mass flow, the work $W_T$ done on the system 
and the heat $Q_T$ absorbed by the system during the time interval $T$
as follows:
\begin{eqnarray}
Z_T&=& \sum_{\lambda=1}^L\mu_\lambda \left\{\omega_T(N_\lambda \otimes{\bf 1}_1 )
-\omega(N_\lambda\otimes{\bf 1}_1)\right\}\cr
W_T&=& \int_0^T dt \ \omega_t\left({\textstyle{d \over dt}}W(t)\otimes {\bf 1}_1\right) \cr
Q_T&=&\{\omega_T(W(T)\otimes {\bf 1}_1+\kappa V)-\omega(W_0\otimes {\bf 1}_1+\kappa V)\}-W_T-Z_T
\nonumber
\end{eqnarray}
Then, one has the following results:
\begin{quote}
\hskip -15pt {\it Proposition 5:\ Stepwise perturbation}\hfil \break 
Suppose $W(t)=W_f$ ($t\ge t_0 >0)$. 
Let ${\alpha}_t^{(f)}$ be the evolution defined by
$$
{d{\alpha}^{(f)}_t(A)\over dt} ={\alpha}^{(f)}_t\left(
{\tilde\delta}_1(A)+i[ W_f\otimes {\bf 1}_1+\kappa V,A]\right) \ ,
\quad {\alpha}^{(f)}_t(A)|_{t=0}=A \ ,
\quad (A\in D({\tilde \delta}_1) )
\ ,
$$
$\omega_f$ be a KMS state with respect to 
${\hat \sigma}_s^{(f)} \equiv \alpha_s^{(f)} \alpha_{-s {\vec \mu}}$ at $\beta$,
and $({\cal H}_f,\pi_f,\Omega_f)$ be its GNS representation\footnote{Namely, there exist
a Hilbert space ${\cal H}_f$ and a *-morphism $\pi_f$ from ${\cal F}$ into the algebra of
all bounded operators on ${\cal H}_f$ and a unit vector $\Omega_f\in {\cal H}_f$, such that 
the closure of $\pi_f({\cal F})\Omega_f$ is equal to ${\cal H}_f$.
}.
Then, if the initial state $\omega$ is the unique KMS state and a self-adjoint operator $L_f$ 
on ${\cal H}_f$ defined
by $e^{iL_f t}\pi_f(A)\Omega_f\equiv \pi_f({\alpha}^{(f)}_t(A))\Omega_f$ $(A\in {\cal F})$, which
will be called the Liouvillian of $\alpha_t^{(f)}$, 
has a simple eigenvalue at zero and the absolute continuous spectrum, 
one has
\begin{eqnarray}
\lim_{\kappa \to 0}\lim_{T\to +\infty}
\beta Q_T&=&S(\rho_f)-S(\rho_i)-S(\rho_f|\rho_{t_0})
\nonumber
\end{eqnarray}
where $S(\rho_\nu)=-{\rm Tr}\{\rho_\nu \ln \rho_\nu\}$ 
($\nu=i,f$) is the von Neumann entropy of the density matrix $\rho_\nu$, 
$S(\rho_f|\rho_{t_0})={\rm Tr}\{\rho_{t_0}(\ln \rho_{t_0}-\ln \rho_f)\}
\ge 0$
is the relative entropy between density matrices $\rho_f$ and $\rho_{t_0}$, and 
$\rho_i=\exp(-\beta(W_0-\sum_\lambda \mu_\lambda N_\lambda))/\Xi_i$,
$\rho_{t_0}=u_{t_0}^*\rho_i u_{t_0}$,
$\rho_f=\exp(-\beta(W_f-\sum_\lambda \mu_\lambda N_\lambda))/\Xi_f$, 
with $\Xi_i$, $\Xi_f$ the grand partition functions.
The unitary element $u_t\in {\cal F}_S$ is the solution of
${d\over dt}u_t = iu_tW(t)$, $u_t|_{t=0}={\bf 1}_S$.
\end{quote}

\medskip

\begin{quote}
\hskip -15 pt{\it Proposition 6: \ Staircase perturbation}
\hfil \break
Suppose that $T=\sum_{j=1}^N
T_j$ and the interaction $W(t)$ has a staircase form:
$W(t)=W_0+(j-1+\varphi(t-{\widetilde T}_{j-1}))(W_f-W_0)/N$ for 
${\widetilde T}_{j-1}\le t \le {\widetilde T}_j$ 
where ${\widetilde T}_j=\sum_{k=1}^j T_k, 
({\widetilde T}_0\equiv 0$) and
$\varphi(t)$ is a twice continuously differentiable real-valued function
with $\varphi(0)=0$, $\varphi(t)=1$ for $t\ge t_0$.
Define $W_j\equiv~W_0+j(W_f-W_0)/N$ and introduce the group
${\alpha}_t^{(j)}$ by
\begin{eqnarray}
{d{\alpha}^{(j)}_t(A)\over dt}\! \! &=&\! \! 
{\alpha}^{(j)}_t\left(
{\tilde\delta}_1(A)+i[ W_j\otimes {\bf 1}_1+\kappa V,A]\right)
\ , 
\quad 
{\alpha}^{(j)}_t(A)|_{t=0}=A \ ,
\quad
(^\forall A\in D({\tilde \delta}_1)) \ .
\nonumber
\end{eqnarray}
Let $\omega_j$ be a KMS state with respect to 
${\hat \sigma}_s^{(j)} \equiv \alpha_s^{(j)}\alpha_{-s {\vec \mu}}$ at $\beta$,
and $({\cal H}_j,\pi_j,\Omega_j)$ be its GNS representation.
Then, if the initial state $\omega$ is the unique KMS state and the Liouvillian $L_j$ 
defined
by $e^{iL_j t}\pi_j(A)\Omega_j\equiv \pi_j({\alpha}^{(j)}_t(A))\Omega_j$ $(A\in {\cal F})$
has a simple eigenvalue at zero and the absolute continuous spectrum, 
one has
\begin{eqnarray}
\lim_{\kappa\to 0} \lim_{T_1\to +\infty}
\cdots \lim_{T_N\to +\infty}
\beta Q_T&=&S(\rho_f)-S(\rho_i) +{\rm O}\left({1\over N}\right)
\nonumber
\end{eqnarray}
\end{quote}

\medskip

\noindent{\bf NB 7}: \
Because of the return-to-equilibrium property, the final state 
(and every intermediate state in Proposition 6) of the whole system is an 
equlibrium state. The limit of $\kappa \to 0$ implies that the coupling between 
the system and the reservoir is negligibly small. 
Thus, the processes treated in Proposition 5
and Proposition 6 precisely
correspond to those in the classical thermodynamics~\cite{Thermody}. 
The Clausius inequality follows from Proposition 5:  $\lim_{\kappa\to 0} \lim_{T\to +\infty}
\beta Q_T\le S(\rho_f)-S(\rho_i)$ 
because of the positivity of the relative entropy.
On the other hand, Proposition 6 implies that, if the whole system changes 
very slowly  ($N\gg 1$) so that the whole system is in equlibrium at every instant, the
Clausius equality holds.
Thus, the thermodynamic entropy is given by the von Neumann entropy,
as expected,  and the process described in Proposition 6 is nothing but a quasistatic process.
Note that, as $\kappa V$ is responsible for the equlibration, the weak 
coupling limit $\kappa \to 0$ should be taken {\it after} the long term limits.

\medskip

\noindent{\bf NB 8}: \ The two propositions deal with the entropy change of a subsystem 
in contrast to the previous 
works~\cite{Ruelle1,JaksicPillet2,Ojima1,Ojima2,JaksicPillet1,RuelleEnPro,FrolichEQ}, 
which have discussed the entropy production of the whole system. 
In this respect, the present work shares a common interest with that by Maes and 
Tasaki\cite{MaesTasaki}, who derived a relation equivalent to the Clausius inequality 
from dynamics for a class of finite classical systems.
The difference between the present setting and that of Ref.~\cite{MaesTasaki} lies in 
the fact that we deal with an infinite-degree-of-freedom reservoir in order to have 
equilibrium states as the final state of Proposition 5 and as the
intermediate states of Proposition 6.

\medskip

\noindent{\bf NB 9}: \ 
Fr\" ohlich, Merkli, Schwarz, and Ueltschi\cite{FrolichEQ} have discussed the Clausius 
equality for `adiabatic' processes where the difference between the true and the 
reference states\footnote{The reference state is defined as an equilibrium
state with instantaneous parameter values.} 
restricted to a certain subsystem is small for all times.
However, conditions on the dynamics which realize the `adiabatic' processes are
not given. Instead, Proposition 6 gives a concrete example of quasi-static processes,
which is similar to the one studied for the classical stochastic systems 
by Sekimoto~\cite{Sekimoto}.

\medskip

\noindent{\bf Proof of Proposition 5:} The heat is rewritten and, then, 
the assertions are shown.

\noindent{\it Expression of heat}: \
Let $\Gamma_t$ be solutions of 
${d\over dt}\Gamma_t= i \Gamma_t {\tilde\tau}^{(1)}_t(W(t)\otimes {\bf 1}_1+\kappa V) $, \ 
$\Gamma_t|_{t=0}={\bf 1}$, where ${\bf 1}\equiv {\bf 1}_S\otimes {\bf 1}_1$, then
$\tau^W_t(A)=\Gamma_t{\tilde \tau}_t^{(1)}(A)\Gamma_t^*$ and 
\begin{eqnarray}
&&-i{d\over dt} \Bigl(({\tilde\delta}_1- \sum_{\lambda=1}^L\mu_\lambda 
{\tilde g}_\lambda^{(1)})(\Gamma_t)\Gamma_t^* \Bigr)
=
({\tilde\delta}_1-\sum_{\lambda=1}^L\mu_\lambda  {\tilde g}_\lambda^{(1)})(
\Gamma_t {\tilde\tau}^{(1)}_t(W(t)\otimes {\bf 1}_1+\kappa V))\Gamma_t^* \cr
&&\mskip 20mu
- ({\tilde\delta}_1-\sum_{\lambda=1}^L\mu_\lambda 
{\tilde g}_\lambda^{(1)} )(\Gamma_t) 
{\tilde\tau}^{(1)}_t(W(t)\otimes {\bf 1}_1+\kappa V)\Gamma_t^* 
=
\tau_t^W \Bigl(({\tilde\delta}_1-\sum_{\lambda=1}^L \mu_\lambda 
{\tilde g}_\lambda^{(1)})(\kappa V)\Bigr)
\label{Heat000}
\end{eqnarray}
where we have used $({\tilde\delta}_1-\sum_{\lambda=1}^L\mu_\lambda 
{\tilde g}_\lambda^{(1)})(AB)
=({\tilde\delta}_1-\sum_{\lambda=1}^L\mu_\lambda 
{\tilde g}_\lambda^{(1)})(A)B+A
({\tilde\delta}_1-\sum_{\lambda=1}^L\mu_\lambda 
{\tilde g}_\lambda^{(1)})(B)$,
${\tilde \delta}_1(W(t)\otimes {\bf 1}_1)=0$ and
${\tilde g}_\lambda^{(1)}(W(t)\otimes {\bf 1}_1)=0$. 
Reminding
\begin{eqnarray}
{d\over dt}\tau_t^W(N_\lambda\otimes {\bf 1}_1 ) 
&=&\tau_t^W\Bigl
({\tilde\delta}_1(N_\lambda\otimes {\bf 1}_1 )+
i[W(t)\otimes{\bf 1}_1+\kappa V,N_\lambda\otimes {\bf 1}_1 ]\Bigr)\cr
&=& \tau_t^W\Bigl(
i[\kappa V,N_\lambda\otimes {\bf 1}_1 ]\Bigr) 
=\tau_t^W({\tilde g}_\lambda^{(1)}(\kappa V))
\ , \\
{d\over dt} \tau_t^W(W(t)\otimes {\bf 1}_1+\kappa V)&=&
\tau_t^W\Bigl({\tilde\delta}_1(W(t)\otimes {\bf 1}_1+\kappa V)+
i[W(t)\otimes{\bf 1}_1+\kappa V,W(t)\otimes {\bf 1}_1+\kappa V]\Bigr)
\cr
&&+\tau_t^W\Bigl({dW(t)\over dt}\otimes {\bf 1}_1\Bigr)
=\tau_t^W\Bigl({\tilde\delta}_1(\kappa V)
\Bigr)
+\tau_t^W\Bigl({dW(t)\over dt}\otimes {\bf 1}_1\Bigr) \ ,
\end{eqnarray}
one obtains

\begin{eqnarray}
&&-i{d\over dt} \left(({\tilde\delta}_1- \sum_{\lambda=1}^L\mu_\lambda 
{\tilde g}_\lambda^{(1)})(\Gamma_t)\Gamma_t^* \right)\cr
&&~=
{d \tau_t^W(W(t)\otimes{\bf 1}_1+\kappa V) \over dt}
-\tau_t^W\left({dW(t)\over dt}\otimes {\bf 1}_1\right)
-\sum_{\lambda=1}^L\mu_\lambda
{d \tau_t^W(N_\lambda\otimes {\bf 1}_1)\over dt}
\label{Heat1}
\end{eqnarray}
and, thus,
\begin{eqnarray}
Q_T&\equiv&
\omega(\tau_T^W(W(T) \otimes {\bf 1}_1 +\kappa V))
-\omega(W(0)\otimes {\bf 1}_1+\kappa V)\cr
&&-\int_0^T ds\omega\left(\tau_s^W\left({dW(s)\over ds}\otimes {\bf 1}_1\right)\right)
-\sum_{\lambda=1}^L\mu_\lambda
\left(\omega(\tau_T^W(N_\lambda\otimes {\bf 1}_1) )-\omega(N_\lambda\otimes {\bf 1}_1 )\right)\cr
&=&
-i\omega\left(
({\tilde\delta}_1- \sum_{\lambda=1}^L\mu_\lambda 
{\tilde g}_\lambda^{(1)})(\Gamma_T)\Gamma_T^* \right) \ .
\label{HEAT0}
\end{eqnarray}

\noindent{\it Proof of the assertions}: \
Let $\Gamma_t^{(f)}$ be the solution of 
${d\over dt}\Gamma_t^{(f)}= i\Gamma_t^{(f)} 
{\tilde\tau}^{(1)}_t(W_f\otimes {\bf 1}_1
+\kappa V)$, \ $\Gamma_t^{(f)}|_{t=0}={\bf 1}$, then, as in the derivation of
(\ref{Heat1}), one obtains
$\alpha_t^{(f)}(A)=\Gamma_t^{(f)}{\tilde \tau}_t^{(1)}(A)\Gamma_t^{(f)*}$,
\begin{eqnarray}
&&-i{d\over dt}\left\{({\tilde\delta}_1- \sum_{\lambda=1}^L\mu_\lambda 
{\tilde g}_\lambda^{(1)})(\Gamma_t^{(f)})\Gamma_t^{(f)*} \right\}
=
\alpha_t^{(f)} \left(({\tilde\delta}_1-\sum_{\lambda=1}^L \mu_\lambda {\tilde g}_\lambda^{(1)})(\kappa V)
\right)
\cr
&&=
{d\over dt}\alpha_t^{(f)}\left( W_f\otimes{\bf 1}_1 +\kappa V -\sum_{\lambda=1}^L
\mu_\lambda N_\lambda\otimes {\bf 1}_1  \right)
\nonumber
\end{eqnarray}
and, thus,
\begin{eqnarray}
-i ({\tilde\delta}_1- \sum_{\lambda=1}^L\mu_\lambda 
{\tilde g}_\lambda^{(1)})(\Gamma_t^{(f)})\Gamma_t^{(f)*} 
&=&
\alpha_t^{(f)}( (W_f-\sum_{\lambda=1}^L
\mu_\lambda N_\lambda )\otimes{\bf 1}_1+\kappa V ) -(W_f-\sum_{\lambda=1}^L
\mu_\lambda N_\lambda)\otimes{\bf 1}_1-\kappa V
\ . \nonumber
\end{eqnarray}
Then, because of $\Gamma_t= \Gamma_{t_0}{\tilde \tau}_{t_0}^{(1)}\left(\Gamma_{t-t_0}^{(f)}\right)$
for $t\ge t_0$, the heat flow to the reservoir is rewritten as
\begin{eqnarray}
&&Q_T = 
-i\omega\left(\Gamma_{t_0}{\tilde \tau}_{t_0}^{(1)}\left(
({\tilde\delta}_1- \sum_{\lambda=1}^L\mu_\lambda 
{\tilde g}_\lambda^{(1)})(\Gamma_{T-t_0}^{(f)})
\Gamma_{T-t_0}^{(f)*}\right)\Gamma_{t_0}^*\right)
-i\omega\left(
({\tilde\delta}_1- \sum_{\lambda=1}^L\mu_\lambda 
{\tilde g}_\lambda^{(1)})(\Gamma_{t_0})
\Gamma_{t_0}^*\right) \cr
&&~~~~= 
\omega\left( \Gamma_{-t_0}^{(i)} {\bar\Gamma}_{t_0}
\alpha_{T-t_0}^{(f)}( (W_f-\sum_{\lambda=1}^L
\mu_\lambda N_\lambda)
\otimes{\bf 1}_1+\kappa V )
{\bar\Gamma}_{t_0}^* \Gamma_{-t_0}^{(i)*}\right)
\cr
&&~~~~~-\omega\left(\Gamma_{t_0}
{\tilde \tau}_{t_0}^{(1)}( (W_f-\sum_{\lambda=1}^L
\mu_\lambda N_\lambda)\otimes{\bf 1}_1)
\Gamma_{t_0}^* \right) 
-\omega\left( \Gamma_{t_0}
{\tilde \tau}_{t_0}^{(1)}(\kappa V)
\Gamma_{t_0}^* \right) \cr
&&~~~~~
-i\omega\left(
({\tilde\delta}_1- \sum_{\lambda=1}^L\mu_\lambda 
{\tilde g}_\lambda^{(1)})(\Gamma_{t_0})
\Gamma_{t_0}^*\right) \ ,
\end{eqnarray}
where a unitary element 
$\Gamma_t^{(i)}$ is the solution of 
${d\over dt}\Gamma_t^{(i)}= i\Gamma_t^{(i)} 
{\tilde\tau}^{(1)}_t(W_0\otimes {\bf 1}_1
+\kappa V)$ with \ $\Gamma_t^{(i)}|_{t=0}={\bf 1}$, 
${\bar{\Gamma}}_{t_0}={\tilde\tau}_{t_0}^{(1)-1}(\Gamma_{t_0})$ is again unitary, 
and, in deriving the first two terms,
we have used a relation $\omega( {\tilde\tau}_t^{(1)}(A)) =\omega(\alpha_t^{(i)}\left({\tilde\tau}_{-t}^{(1)}(A)\right)) 
=\omega(\Gamma_{-t}^{(i)}A \Gamma_{-t}^{(i)*})$
resulting from the $\alpha_t^{(i)}$-invariance of $\omega$.

Let us begin with the evaluation of the $T$-independent terms. 
Since (\ref{Heat000}) gives
\begin{eqnarray}
-i({\tilde\delta}_1- \sum_{\lambda=1}^L\mu_\lambda 
{\tilde g}_\lambda^{(1)})
(\Gamma_{t_0}) \Gamma_{t_0}^*
&=& \kappa \int_0^{t_0}ds \tau_{s}^W\left(
({\tilde\delta}_1-\sum_{\lambda=1}^L \mu_\lambda {\tilde g}_\lambda^{(1)})(V)
\right) \ ,
\end{eqnarray}
the last term is evaluated as
\begin{eqnarray}
&&\left|\omega\left(
({\tilde\delta}_1- \sum_{\lambda=1}^L\mu_\lambda 
{\tilde g}_\lambda^{(1)})(\Gamma_{t_0})
\Gamma_{t_0}^*\right)\right|
\le
\left\Vert
({\tilde\delta}_1- \sum_{\lambda=1}^L\mu_\lambda 
{\tilde g}_\lambda^{(1)})(\Gamma_{t_0})
\Gamma_{t_0}^*\right\Vert
\cr
&&\le
|\kappa| \int_0^{t_0}ds 
\left\Vert \tau_{s}^W\left(
({\tilde\delta}_1-\sum_{\lambda=1}^L 
\mu_\lambda {\tilde g}_\lambda^{(1)})(V)
\right) 
\right\Vert
=|\kappa| t_0  
\left\Vert 
({\tilde\delta}_1-\sum_{\lambda=1}^L \mu_\lambda 
{\tilde g}_\lambda^{(1)})(V)
\right\Vert
\ .
\end{eqnarray}

Since $\Vert\Gamma_{t_0}\Vert=1$ and
$\Vert{\tilde \tau}_{t_0}^{(1)}(V)\Vert=\Vert V\Vert$,
the third term is bounded by $|\kappa| \Vert V\Vert$:
$$
\left|\omega\left( \Gamma_{t_0}
{\tilde \tau}_{t_0}^{(1)}(\kappa V)
\Gamma_{t_0}^* \right)\right|
\le |\kappa| \ \Vert V\Vert \ .
$$

By taking into account the fact
that ${\tilde\tau}_t^{(1)}(W(t)\otimes {\bf 1}_1)=W(t)\otimes {\bf 1}_1$ and
comparing the equations for $\Gamma_t$ and $u_t$, one obtains
\begin{eqnarray}
\Gamma_{t_0}=u_{t_0}\otimes {\bf 1}_1 +\Delta\Gamma_{t_0} \ , \qquad
\Delta\Gamma_{t_0}
\equiv
i\kappa \int_0^{t_0}ds \Gamma_s{\tilde\tau}_s^{(1)}(V)((u_s^*u_{t_0})\otimes{\bf 1}_1)
\ .
\end{eqnarray}
Then, because of ${\tilde \tau}_{t_0}^{(1)}( (W_f-\sum_{\lambda=1}^L\mu_\lambda N_\lambda)\otimes{\bf 1}_1)
=(W_f-\sum_{\lambda=1}^L
\mu_\lambda N_\lambda)\otimes{\bf 1}_1$, the second term is evaluated as
\begin{eqnarray}
&&\left|
\omega\Bigl(\Gamma_{t_0}
{\tilde \tau}_{t_0}^{(1)}( (W_f-\sum_{\lambda=1}^L
\mu_\lambda N_\lambda)\otimes{\bf 1}_1)
\Gamma_{t_0}^* \Bigr) 
-\omega\Bigl(
(u_{t_0}(W_f-\sum_{\lambda=1}^L
\mu_\lambda N_\lambda) u_{t_0}^*)\otimes{\bf 1}_1 
\Bigr)\right| \cr
&&=\left|\omega\Bigl(\Gamma_{t_0}
( (W_f-\sum_{\lambda=1}^L
\mu_\lambda N_\lambda)\otimes{\bf 1}_1)
\Gamma_{t_0}^*\Bigr)-
\omega\Bigl(
(u_{t_0}(W_f-\sum_{\lambda=1}^L
\mu_\lambda N_\lambda) u_{t_0}^*)\otimes{\bf 1}_1 
\Bigr)\right| \cr
&&\le
\left|\omega\Bigl(\Delta\Gamma_{t_0}
( (W_f-\sum_{\lambda=1}^L
\mu_\lambda N_\lambda)\otimes{\bf 1}_1)
\Gamma_{t_0}^*\Bigr)\right|+
\left|\omega\Bigl(
(u_{t_0}(W_f-\sum_{\lambda=1}^L
\mu_\lambda N_\lambda))\otimes{\bf 1}_1 \Delta\Gamma_{t_0}^*
\Bigr)\right|
\cr
&&\le
2 \Vert\Delta\Gamma_{t_0}\Vert \
\Vert(W_f-\sum_{\lambda=1}^L
\mu_\lambda N_\lambda)\otimes{\bf 1}_1\Vert 
\le 2 |\kappa| t_0 \Vert V\Vert
\
\Vert(W_f-\sum_{\lambda=1}^L
\mu_\lambda N_\lambda)\otimes{\bf 1}_1\Vert \ ,
\end{eqnarray}
where we have used 
$\Vert\Delta\Gamma_{t_0}\Vert \le |\kappa| t_0 \Vert V\Vert$.

Now we turn to the first term, which is estimated with the aid of the following lemma:

\noindent{\bf Lemma}: 
\begin{equation}
\lim_{t\to \infty}\omega(A\alpha_t^{(f)}(B)C)=\omega(AC)\omega_f(B)
\end{equation}
\begin{quote}
{\it Proof}: As the KMS state $\omega_f$ is the bounded perturbation of the unique KMS state $\omega$ and 
${\hat \sigma}^{(f)}_s$
commutes with $\alpha_t^{(f)}$ for all $t,s\in {\bf R}$, $\omega_f$ is $\alpha_t^{(f)}$-invarinat and, thus,
$L_f$ is well-defined (cf. Corollary 2.3.17 of Ref.~\cite{Bratteli}). 
Then, because of the assumption on the spectrum of $L_f$, one has, for $t\to \infty$,
$$
(\psi,\pi_f(\alpha_t^{(f)}(B))\Omega_f) -(\psi, \Omega_f) (\Omega_f, \pi_f(B)\Omega_f) 
=({\widetilde\psi},e^{iL_f t}{\widetilde \varphi}_B)
=\int_{-\infty}^\infty d\lambda e^{i\lambda t}{d({\widetilde\psi},{\hat E}(\lambda){\widetilde \varphi}_B)
\over d\lambda} 
\to 0 \ ,
$$
where ${\widetilde\psi}=\psi-(\Omega_f,\psi)\Omega_f$ and ${\widetilde\varphi}_B=\pi_f(B)\Omega_f
-(\Omega_f,\pi_f(B)\Omega_f)\Omega_f$ are absolutely continuous vectors,
$\{{\hat E}(\lambda)\}$ is the spectral family of $L_f$ and the Riemann-Lebesgue lemma is used.
Corollary 5.3.9 of Ref.~\cite{Bratteli} asserts that $\Omega_f$ is separating for 
the bi-commutant $\pi_f({\cal F})^{\prime\prime}$ of $\pi_f({\cal F})$ since
$\omega_f$ is a KMS state. 
Then, 
as in the proof of Theorem 4.3.23 of Ref.~\cite{Bratteli}, 
one concludes 
$$
\lim_{t\to \infty}(\psi,\pi_f(\alpha_t^{(f)}(B))\varphi) =(\psi, \varphi) (\Omega_f, \pi_f(B)\Omega_f) 
=(\psi, \varphi) \omega_f(B) \ ,
$$
from the above formula. 
As $\omega$ is a bounded perturbation of $\omega_f$, it is expressed by a vector $\Omega_0\in {\cal H}_f$:
$\omega(A)= (\Omega_0,\pi_f(A)\Omega_0)/\Vert \Omega_0 \Vert^2$ (cf. Corollary 5.4.5 of Ref.~\cite{Bratteli})
and, thus, one obtains the desired result:
\begin{eqnarray}
\lim_{t\to \infty}\omega(A\alpha_t^{(f)}(B)C) &=& \lim_{t\to \infty} (\Omega_0,\pi_f(A)\pi_f(\alpha_t^{(f)}(B))
\pi_f(C)\Omega_0)/\Vert \Omega_0 \Vert^2 \cr
&=&\omega_f(B) (\Omega_0,\pi_f(A)\pi_f(C)\Omega_0)/\Vert \Omega_0 \Vert^2
=
\omega(AC) \omega_f(B) \ . 
\nonumber
\end{eqnarray}
\end{quote}

\medskip

\noindent
From this lemma, one immediately obtains
\begin{eqnarray}
&&\lim_{T\to +\infty}
\omega\Bigl( \Gamma_{-t_0}^{(i)} {\bar\Gamma}_{t_0}
\alpha_{T-t_0}^{(f)}\Bigl( (W_f-\sum_{\lambda=1}^L
\mu_\lambda N_\lambda)
\otimes{\bf 1}_1
+\kappa V \Bigr)
{\bar\Gamma}_{t_0}^* \Gamma_{-t_0}^{(i)*} \Bigr)
\cr
&&=
\omega_f\Bigl( (W_f-\sum_{\lambda=1}^L
\mu_\lambda N_\lambda)
\otimes{\bf 1}_1+\kappa V 
\Bigr)
\  
\omega\Bigl( \Gamma_{-t_0}^{(i)} {\bar\Gamma}_{t_0}
{\bar\Gamma}_{t_0}^* \Gamma_{-t_0}^{(i)*}\Bigr)
\cr
&&=\omega_f\Bigl(
(W_f-\sum_{\lambda=1}^L
\mu_\lambda N_\lambda)
\otimes{\bf 1}_1
\Bigr) 
+\omega_f(\kappa V) \ .
\end{eqnarray}
Moreover, $|\omega_f(\kappa V)|\le |\kappa| \Vert V\Vert$.

In short, we have shown
\begin{eqnarray}
\lim_{T\to +\infty}Q_T &=& 
\omega_f\Bigl(
(W_f-\sum_{\lambda=1}^L
\mu_\lambda N_\lambda)
\otimes{\bf 1}_1
\Bigr)
-
\omega\Bigl((u_{t_0}
(W_f-\sum_{\lambda=1}^L
\mu_\lambda N_\lambda)u_{t_0}^*)
\otimes{\bf 1}_1\Bigr)
+{\rm O}(|\kappa|)
\nonumber
\end{eqnarray}
Remind that
$\omega$ and $\omega_f$ are $\kappa V$-perturbed KMS states of the product states
$\rho_i\otimes \omega_{GC}$ and $\rho_f\otimes \omega_{GC}$, respectively, where $\omega_{GC}$ is
the reservoir grand canonical state
(a KMS state at $\beta$ with respect to ${\tilde\tau}_t^{(1)}{\tilde\alpha}^{(1)}_{-{\vec \mu} t}$).
Then, 
the stability of KMS states (Theorem 5.4.4 of Ref.~\cite{Bratteli})
leads to $\lim_{\kappa\to 0}\omega_f(A\otimes{\bf 1}_1)={\rm Tr}(\rho_f A)$
and $\lim_{\kappa\to 0}\omega(A\otimes{\bf 1}_1)={\rm Tr}(\rho_i A)$ ($^\forall A\in {\cal F}_S$,
see also Ref.~\cite{Ohya}) and, hence,
\begin{eqnarray}
&&\lim_{\kappa\to 0}\lim_{T\to +\infty}\beta Q_T = 
\beta \ {\rm Tr}\rho_f\Bigl(W_f-\sum_{\lambda=1}^L
\mu_\lambda N_\lambda\Bigr)
-
\beta \ {\rm Tr}\rho_i\Bigl(u_{t_0}(W_f-\sum_{\lambda=1}^L
\mu_\lambda N_\lambda)u_{t_0}^*\Bigr)\cr
&&= 
-{\rm Tr}( \rho_f \ln \rho_f)
+{\rm Tr}\rho_i(u_{t_0}(\ln \rho_f)u_{t_0}^*)
=
-{\rm Tr}( \rho_f \ln \rho_f)
+
{\rm Tr}(\rho_{t_0}(\ln \rho_f-\ln \rho_{t_0}))
+{\rm Tr}( \rho_i \ln \rho_i)\cr
&&=S(\rho_f)-S(\rho_i)-S(\rho_f|\rho_{t_0})
\ .
\nonumber
\end{eqnarray}

\medskip

\noindent {\bf Proof of Proposition 6:} The formula (\ref{HEAT0}) is still valid:
\begin{eqnarray}
Q_T&=&-i \omega\Bigl(({\tilde \delta}_1-\sum_\lambda \mu_\lambda{\tilde g}_\lambda^{(1)})(\Gamma_T)\Gamma_T^*\Bigr) \cr
&=&-i\sum_{j=1}^N
\left[
\omega\Bigl(({\tilde \delta}_1-\sum_\lambda \mu_\lambda{\tilde g}_\lambda^{(1)})(\Gamma_{{\tilde T}_j})\Gamma_{{\tilde T}_j}^*\Bigr) 
-\omega\Bigl(({\tilde \delta}_1-\sum_\lambda \mu_\lambda{\tilde g}_\lambda^{(1)})
(\Gamma_{{\tilde T}_{j-1}})\Gamma_{{\tilde T}_{j-1}}^*\Bigr) 
\right] \ .
\end{eqnarray}
Let ${\widetilde \Gamma}^{(j)}_t$ be the solution of ${\widetilde \Gamma}^{(j)}_0={\bf 1}$, 
${d\over dt} {\widetilde \Gamma}^{(j)}_t = i{\widetilde \Gamma}^{(j)}_t{\tilde\tau}_t^{(1)}(W_j\otimes{\bf 1}_1+\kappa V)$
and
${\widetilde \Gamma}^{\prime\prime(j)}_t$ be the solution of ${\widetilde \Gamma}^{\prime\prime(j)}_0={\bf 1}$,
${d\over dt} {\widetilde \Gamma}^{\prime\prime (j)}_t = i{\widetilde \Gamma}^{\prime\prime(j)}_t{\tilde\tau}_t^{(1)}
((W_{j-1}+\varphi(t)\Delta W)\otimes{\bf 1}_1 + \kappa V)$ 
with 
\break
$\Delta W=(W_f-W_0)/N$,
then, by comparing these equations with that of $\Gamma_t$, one has
${\widetilde\Gamma}_t^{(j)} {\widetilde\tau}_t^{(1)}(A){\widetilde\Gamma}_t^{(j)*}
=\alpha_t^{(j)}(A)$,
$\omega({\widetilde\tau}_t^{(1)}(A))=\omega(\alpha_{-t}^{(0)}({\widetilde\tau}_t^{(1)}(A)))
=\omega( {\widetilde\Gamma}_{-t}^{(0)}A{\widetilde\Gamma}_{-t}^{(0)*})$,  \break
$\Gamma_{{\tilde T}_j}=\Gamma_{{\tilde T}_{j-1}+t_0}{\tilde\tau}_{{\tilde T}_{j-1}+t_0}^{(1)}
\left({\widetilde \Gamma}^{(j)}_{T_j-t_0}\right)$
and
$\Gamma_{{\tilde T}_{j-1}+t_0}=\Gamma_{{\tilde T}_{j-1}} {\tilde\tau}_{{\tilde T}_{j-1}}^{(1)}
\left( {\widetilde \Gamma}^{\prime\prime(j)}_{t_0} \right)$. \
Moreover,
$$
-i({\tilde \delta}_1-\sum_\lambda \mu_\lambda {\tilde g}_\lambda^{(1)})( {\widetilde \Gamma}_t^{(j)})
{\widetilde \Gamma}_t^{(j)*}
=\alpha_t^{(j)}( (W_j-\sum_\lambda \mu_\lambda N_\lambda)\otimes {\bf 1}_1+\kappa V)
-(W_j-\sum_\lambda \mu_\lambda N_\lambda)\otimes {\bf 1}_1-\kappa V
$$
Hence, we have
\begin{eqnarray}
&&-i\omega\Bigl(({\tilde \delta}_1-\sum_\lambda \mu_\lambda{\tilde g}_\lambda^{(1)})
(\Gamma_{{\tilde T}_j})\Gamma_{{\tilde T}_j}^*\Bigr) 
+i\omega\Bigl(({\tilde \delta}_1-\sum_\lambda \mu_\lambda{\tilde g}_\lambda^{(1)})
(\Gamma_{{\tilde T}_{j-1}})\Gamma_{{\tilde T}_{j-1}}^*\Bigr) 
\cr
&&~=
-i\omega\Bigl( \Gamma_{{\tilde T}_{j-1}+t_0} {\tilde\tau}_{{\tilde T}_{j-1}+t_0}^{(1)}\Bigl(
({\tilde \delta}_1-\sum_\lambda \mu_\lambda{\tilde g}_\lambda^{(1)})
({\widetilde \Gamma}^{(j)}_{T_j-t_0} ){\widetilde \Gamma}^{(j)*}_{T_j-t_0} 
\Bigr)\Gamma_{{\tilde T}_{j-1}+t_0}^* \Bigr) 
\cr
&&~~~-i\omega\Bigl(({\tilde \delta}_1-\sum_\lambda \mu_\lambda{\tilde g}_\lambda^{(1)})(\Gamma_{{\tilde T}_{j-1}+t_0} )
\Gamma_{{\tilde T}_{j-1}+t_0}^*\Bigr) 
+i\omega\Bigl(({\tilde \delta}_1-\sum_\lambda \mu_\lambda{\tilde g}_\lambda^{(1)})
(\Gamma_{{\tilde T}_{j-1}})\Gamma_{{\tilde T}_{j-1}}^*\Bigr) 
\cr
&&~=
-i\omega\Bigl( \Gamma_{{\tilde T}_{j-1}+t_0} {\tilde\tau}_{{\tilde T}_{j-1}+t_0}^{(1)}\Bigl(
({\tilde \delta}_1-\sum_\lambda \mu_\lambda{\tilde g}_\lambda^{(1)})
({\widetilde \Gamma}^{(j)}_{T_j-t_0} ) 
{\widetilde \Gamma}^{(j)*}_{T_j-t_0} \Bigr)
\Gamma_{{\tilde T}_{j-1}+t_0}^* \Bigr) \cr
&&~~~-i\omega\Bigl(\Gamma_{{\tilde T}_{j-1}} {\tilde\tau}_{{\tilde T}_{j-1}}^{(1)}
\Bigl(({\tilde \delta}_1-\sum_\lambda \mu_\lambda{\tilde g}_\lambda^{(1)})( {\widetilde \Gamma}^{\prime\prime(j)}_{t_0})
\Bigr)
\Gamma_{{\tilde T}_{j-1}+t_0}^*\Bigr) 
\cr
&&~=
\omega\Bigl( \Gamma_{{\tilde T}_{j-1}+t_0}{\tilde\tau}_{{\tilde T}_{j-1}+t_0}^{(1)}\Bigl(
\alpha_{T_j-t_0}^{(j)}\Bigl(
(W_j-\sum_\lambda \mu_\lambda N_\lambda)\otimes {\bf 1}_1
+\kappa V
\Bigr)\Bigr)
\Gamma_{{\tilde T}_{j-1}+t_0}^* \Bigr) 
\cr
&&
~~~-\omega\Bigl( \Gamma_{{\tilde T}_{j-1}+t_0}
{\tilde\tau}_{{\tilde T}_{j-1}+t_0}^{(1)}\Bigl(
(W_j-\sum_\lambda \mu_\lambda N_\lambda)\otimes {\bf 1}_1
+\kappa V
\Bigr)
\Gamma_{{\tilde T}_{j-1}+t_0}^* \Bigr)
\cr
&&~~~-i\omega\Bigl(\Gamma_{{\tilde T}_{j-1}} {\tilde\tau}_{{\tilde T}_{j-1}}^{(1)}
\Bigl(({\tilde \delta}_1-\sum_\lambda \mu_\lambda{\tilde g}_\lambda^{(1)})( {\widetilde \Gamma}^{\prime\prime(j)}_{t_0})
{\widetilde \Gamma}^{\prime\prime(j)*}_{t_0}
\Bigr)
\Gamma_{{\tilde T}_{j-1}}^*\Bigr) 
\cr
&&~=
\omega\Bigl( {\widetilde \Gamma}_{-{\tilde T}_{j-1}-t_0}^{(0)}
{\bar\Gamma}_{{\tilde T}_{j-1}+t_0}
\Bigl\{
\alpha_{T_j-t_0}^{(j)}\Bigl(
(W_j-\sum_\lambda \mu_\lambda N_\lambda)\otimes {\bf 1}_1
+\kappa V
\Bigr)\Bigr\}
{\bar\Gamma}_{{\tilde T}_{j-1}+t_0}^* 
{\widetilde \Gamma}_{-{\tilde T}_{j-1}-t_0}^{(0)*}
\Bigr) 
\cr
&&
~~~- \omega\Bigl( \Gamma_{{\tilde T}_{j-1}}
{\tilde\tau}_{{\tilde T}_{j-1}}^{(1)}\Bigl(
{\widetilde \Gamma}^{\prime\prime(j)}_{t_0}
{\tilde\tau}_{t_0}^{(1)}\Bigl(
(W_j-\sum_\lambda \mu_\lambda N_\lambda)\otimes {\bf 1}_1
+\kappa V
\Bigr)
{\widetilde \Gamma}^{\prime\prime(j)*}_{t_0}
\Bigr)
\Gamma_{{\tilde T}_{j-1}}^* \Bigr)
\cr
&&~~~-i \omega\Bigl(\Gamma_{{\tilde T}_{j-1}} {\tilde\tau}_{{\tilde T}_{j-1}}^{(1)}
\Bigl(({\tilde \delta}_1-\sum_\lambda \mu_\lambda{\tilde g}_\lambda^{(1)})( {\widetilde \Gamma}^{\prime\prime(j)}_{t_0})
{\widetilde \Gamma}^{\prime\prime(j)*}_{t_0}
\Bigr)
\Gamma_{{\tilde T}_{j-1}}^*\Bigr) \ , \nonumber
\end{eqnarray}
where
${\bar\Gamma}_t\equiv{\widetilde\tau}_t^{(1)-1}(\Gamma_t)$.
On the other hand, the states $\omega, \omega_j$ and the evolutions $\alpha_t^{(j)}$ ($j=1,2,\cdots N$) 
satisfy the assumptions necessary for the Lemma, we have
\begin{eqnarray}
&&~\lim_{T_j\to+\infty}
\omega\Bigl( {\widetilde \Gamma}_{-{\tilde T}_{j-1}-t_0}^{(0)}
{\bar\Gamma}_{{\tilde T}_{j-1}+t_0}
\alpha_{T_j-t_0}^{(j)}\Bigl(A
\Bigr)
{\bar\Gamma}_{{\tilde T}_{j-1}+t_0}^* 
{\widetilde \Gamma}_{-{\tilde T}_{j-1}-t_0}^{(0)*}
\Bigr) 
\cr
&&~~~~=
\omega\left( {\widetilde \Gamma}_{-{\tilde T}_{j-1}-t_0}^{(0)}
{\bar\Gamma}_{{\tilde T}_{j-1}+t_0}
{\bar\Gamma}_{{\tilde T}_{j-1}+t_0}^* 
{\widetilde \Gamma}_{-{\tilde T}_{j-1}-t_0}^{(0)*}
\right) 
\omega_j\left(A
\right)
=\omega_j(A) \ , \\
&&\lim_{T_{j-1}\to+\infty}
\omega\left(\Gamma_{{\tilde T}_{j-1}} {\tilde\tau}_{{\tilde T}_{j-1}}^{(1)}
(A)
\Gamma_{{\tilde T}_{j-1}}^*\right) 
=
\lim_{T_{j-1}\to+\infty}
\omega\left( \Gamma_{{\tilde T}_{j-2}+t_0} {\tilde \tau}_{{\tilde T}_{j-2}+t_0}^{(1)}\left(
\alpha_{T_{j-1}-t_0}^{(j)}\left(A
\right)\right)
\Gamma_{{\tilde T}_{j-2}+t_0}^*
\right) 
\cr
&&
~~~~=
\omega\left( {\widetilde \Gamma}_{-{\tilde T}_{j-1}-t_0}^{(0)}
{\bar\Gamma}_{{\tilde T}_{j-2}+t_0} 
{\bar\Gamma}_{{\tilde T}_{j-2}+t_0}^*
{\widetilde \Gamma}_{-{\tilde T}_{j-2}-t_0}^{(0)*}
\right)\omega_{j-1}(A) =\omega_{j-1}(A) \ .
\end{eqnarray}
Therefore, by taking the limits $T_j \to+\infty \ , T_{j-1}\to +\infty$ in this order,
we get 
\newpage
\begin{eqnarray}
&&\lim_{T_{j-1}\to\infty}\lim_{T_j\to\infty}
\Bigl[
-i\omega\Bigl(({\tilde \delta}_1-\sum_\lambda \mu_\lambda{\tilde g}_\lambda^{(1)})
(\Gamma_{{\tilde T}_j})\Gamma_{{\tilde T}_j}^*\Bigr) 
+i\omega\Bigl(({\tilde \delta}_1-\sum_\lambda \mu_\lambda{\tilde g}_\lambda^{(1)})
(\Gamma_{{\tilde T}_{j-1}})\Gamma_{{\tilde T}_{j-1}}^*\Bigr) 
\Bigr]
\cr
&&~=
\omega_j\Bigl(
(W_j-\sum_\lambda \mu_\lambda N_\lambda)\otimes {\bf 1}_1
+\kappa V
\Bigr) 
- \omega_{j-1}\Bigl( 
{\widetilde\Gamma}_{t_0}^{\prime\prime(j)} 
{\widetilde\tau}_{t_0}^{(1)}
\Bigl(
(W_j-\sum_\lambda \mu_\lambda N_\lambda)\otimes {\bf 1}_1
+\kappa V
\Bigr)
{\widetilde\Gamma}_{t_0}^{\prime\prime(j)*}
\Bigr)
\cr
&&~~~-i \omega_{j-1}\Bigl( ({\tilde \delta}_1-\sum_\lambda \mu_\lambda{\tilde g}_\lambda^{(1)})
( {\widetilde \Gamma}^{\prime\prime(j)}_{t_0})
{\widetilde \Gamma}^{\prime\prime(j)*}_{t_0}
\Bigr) \ .
\nonumber
\end{eqnarray}
As before, one has
\begin{eqnarray}
&&\max\left\{|\omega_j(\kappa V)| , 
\Big|\omega_{j-1}\bigl( 
{\widetilde\Gamma}_{t_0}^{\prime\prime(j)} 
{\widetilde\tau}_{t_0}^{(1)}
(\kappa V)
{\widetilde\Gamma}_{t_0}^{\prime\prime(j)*}
\bigr)\Big|\right\}\le |\kappa| \ \Vert V\Vert \ ,
\\
&&\Big|\omega_{j-1}({\tilde\delta}_1-\sum_\lambda \mu_\lambda {\tilde g}_\lambda^{(1)})
({\widetilde\Gamma}_{t_0}^{\prime\prime(j)}) 
{\widetilde\Gamma}_{t_0}^{\prime\prime(j)*}\Big| \le |\kappa| \ t_0\ \Vert 
({\tilde\delta}_1-\sum_\lambda \mu_\lambda {\tilde g}_\lambda^{(1)})
(V)\Vert \ ,
\\
&&
\Big|\omega_{j-1}\Bigl( 
{\widetilde\Gamma}_{t_0}^{\prime\prime(j)} 
{\widetilde\tau}_{t_0}^{(1)}
\Bigl(
(W_j-\sum_\lambda \mu_\lambda N_\lambda)\otimes {\bf 1}_1
\Bigr)
{\widetilde\Gamma}_{t_0}^{\prime\prime(j)*}
\Bigr)
-
\omega_{j-1}\Bigl( 
[{\widetilde u}_{t_0}^{(j)} (W_j-\sum_\lambda \mu_\lambda N_\lambda){\widetilde u}_{t_0}^{(j)*}]\otimes {\bf 1}_1
\Bigr)\Big| 
\cr
&&~~\le 2 |\kappa|\ t_0 \ \left\Vert V \right\Vert
\Big\Vert (W_j-\sum_\lambda \mu_\lambda N_\lambda)\otimes {\bf 1}_1 \Big\Vert \ ,
\end{eqnarray}
where ${\widetilde u}_t^{(j)}$
is the solution of ${\widetilde u}_{0}^{(j)}={\bf 1}_S$,
${d\over dt} {\widetilde u}^{(j)}_t = i{\widetilde u}^{(j)}_t(W_{j-1}+\varphi(t) \Delta W)$.
Moreover, the state $\omega_j$ is a $\kappa V$-perturbed KMS state of the product state
$\rho_j\otimes \omega_{GC}$, where
\break
$\rho_j=\exp(-\beta(W_j-\sum_\lambda \mu_\lambda N_\lambda))/\Xi_j$
with $\Xi_j$ the grand partition function, and, thus, $\lim_{\kappa\to 0}\omega_j(A\otimes{\bf 1}_1)
={\rm Tr}(A\rho_j)$ ($^\forall A\in{\cal F}_S$). 
As a consequence, one gets
\begin{eqnarray}
&&\lim_{\kappa\to 0}\lim_{T_{j-1}\to\infty}\lim_{T_j\to\infty}
\left[
-i\omega\left(({\tilde \delta}_1-\sum_\lambda \mu_\lambda{\tilde g}_\lambda^{(1)})
(\Gamma_{{\tilde T}_j})\Gamma_{{\tilde T}_j}^*\right) 
+i\omega\left(({\tilde \delta}_1-\sum_\lambda \mu_\lambda{\tilde g}_\lambda^{(1)})
(\Gamma_{{\tilde T}_{j-1}})\Gamma_{{\tilde T}_{j-1}}^*\right) 
\right]
\cr
&&~=
{\rm Tr}\Bigl(\rho_j
(W_j-\sum_\lambda \mu_\lambda N_\lambda)
\Bigr) 
- {\rm Tr}\Bigl(\rho_{j-1}
({\widetilde u}_{t_0}^{(j)} (W_j-\sum_\lambda \mu_\lambda N_\lambda){\widetilde u}_{t_0}^{(j)*})
\Bigr)
\cr
&&~= \beta^{-1} \left\{-{\rm Tr}(\rho_j\ln \rho_j)
+{\rm Tr}\Bigl({\widetilde\rho}_{j-1}\ln \rho_j)\right\}
=\beta^{-1} \left\{S(\rho_j)-S(\rho_{j-1})-S(\rho_j|{\widetilde\rho}_{j-1})\right\}
\ .
\nonumber
\end{eqnarray}
where ${\widetilde\rho}_{j-1}= {\widetilde u}_{t_0}^{(j)*}\rho_{j-1}{\widetilde u}_{t_0}^{(j)}$, and,
thus,
\begin{eqnarray}
\lim_{\kappa\to 0} \lim_{T_1\to +\infty} \lim_{T_2\to +\infty} \cdots \lim_{T_N\to +\infty} \beta Q_T
&=& S(\rho_f)-S(\rho_i)-\sum_{j=1}^N S(\rho_j|{\widetilde \rho}_{j-1}) \ ,
\nonumber
\end{eqnarray}
where $\rho_i\equiv \rho_0$ and $\rho_f\equiv \rho_N$. 

Now we investigate the contribution from the relative entropies. Reminding that $N_\lambda$ commutes with $W_0$
and $W_f$, one finds
$$
S(\rho_j|{\widetilde\rho_{j-1})=
\beta\langle \Delta {\widetilde u}_{t_0}^{(j)}W {\widetilde u}_{t_0}^{(j)*}\rangle_{j-1}+\beta( \langle {\widetilde u}_{t_0}^{(j)} W_{j-1}
{\widetilde u}_{t_0}^{(j)*}
\rangle_{j-1}
 -\langle W_{j-1}\rangle_{j-1})
+\ln\langle e^{\beta W_{j-1}}e^{-\beta W_j}\rangle_{j-1}} \ ,
$$ 
where $\langle A \rangle_{j-1}\equiv {\rm Tr}(\rho_{j-1}A)$ ($\forall A\in {\cal F}_S$).
On the other hand, we have 
\begin{eqnarray}
&&{\widetilde u}_{t_0}^{(j)} \Delta W {\widetilde u}_{t_0}^{(j)*} = e^{iW_{j-1}t_0}\Delta W e^{-iW_{j-1}t_0}
+ e^{iW_{j-1}t_0}\Delta W \ \Delta{\widetilde u}_{t_0}^{(j)*} + \Delta{\widetilde u}_{t_0}^{(j)} \
\Delta W {\widetilde u}_{t_0}^{(j)*} \ ,
\cr
&&{\widetilde u}_{t_0}^{(j)} W_{j-1} {\widetilde u}_{t_0}^{(j)*} - W_{j-1}=i\int_0^{t_0}ds \varphi(s)
e^{iW_{j-1}s}[\Delta W,W_{j-1}] e^{-iW_{j-1}s} \cr
&&\mskip 120 mu +i\int_0^{t_0}ds \varphi(s) \left\{
e^{iW_{j-1}s}[\Delta W, W_{j-1}] \ \Delta{\widetilde u}_{s}^{(j)*} + \Delta{\widetilde u}_{s}^{(j)} \
[\Delta W , W_{j-1}] {\widetilde u}_{s}^{(j)*}
\right\} \ ,
\cr
&&
e^{\beta W_{j-1}}e^{-\beta W_j}={\bf 1}_S -\int_0^\beta d\tau e^{\tau W_{j-1}} \Delta W e^{-\tau W_{j-1}}
\cr
&&\mskip 120 mu 
+\int_0^\beta d\tau \int_0^\tau d\tau' e^{\tau W_{j-1}} \Delta W e^{-(\tau-\tau') W_{j-1}} \Delta W e^{-\tau' W_j}
\ ,
\nonumber
\end{eqnarray}
where $\Delta{\widetilde u}_{t_0}^{(j)}=i\int_0^{t_0}ds \varphi(s) {\widetilde u}_s^{(j)} \Delta W 
e^{iW_{j-1}(t_0-s)}$, and, since $\rho_{j-1}$ commutes with $W_{j-1}$ and
$\Vert W_j\Vert \le \Vert W_0\Vert +j/N\Vert W_f-W_0\Vert \le \Vert W_0\Vert +\Vert W_f-W_0\Vert \equiv K$,
\begin{eqnarray}
&&\left| \langle {\widetilde u}_{t_0}^{(j)} \Delta W {\widetilde u}_{t_0}^{(j)*} \rangle_{j-1}
- \langle \Delta W \rangle_{j-1}\right| \le 2 \Vert \Delta W\Vert^2 \int_0^{t_0}ds|\varphi(s)|
\ ,
\cr
&&\left| \langle {\widetilde u}_{t_0}^{(j)} W_{j-1} {\widetilde u}_{t_0}^{(j)*} \rangle_{j-1} - 
\langle W_{j-1}\rangle_{j-1}\right|\le
4K \Vert \Delta W\Vert^2 \int_0^{t_0}ds\int_0^s ds' |\varphi(s)\varphi(s')|
\ ,
\cr
&&
|\langle e^{\beta W_{j-1}}e^{-\beta W_j}\rangle_{j-1} -1
+
\beta \langle \Delta W \rangle_{j-1}|
\le \Vert \Delta W\Vert^2
\int_0^\beta d\tau \int_0^\tau d\tau' e^{K(\tau+\tau'+|\tau-\tau'|)}
\ ,
\nonumber
\end{eqnarray}
where we have used $\langle [\Delta W,W_{j-1}] \rangle_{j-1}=0$. Therefore, if $N$ is large
enough, we obtain $|S(\rho_j|{\widetilde\rho_{j-1}})|\le K'\Vert \Delta W\Vert ^2= K'\Vert W_f-W_0\Vert ^2/N^2$
with some positive constant $K'$, and
the desired result:
\begin{eqnarray}
\left|\lim_{\kappa\to 0} \lim_{T_1\to +\infty} \lim_{T_2\to +\infty} \cdots \lim_{T_N\to +\infty} \beta Q_T
- \{S(\rho_f)-S(\rho_i)\}\right| \le {K'\Vert W_f-W_0\Vert ^2\over N}
\ .
\nonumber
\end{eqnarray}

\section{Conclusion}\label{Sec:5}

In the first half of this article, we have shown that, if the evolution is $L^1$-asymptotic abelian 
and the M\o ller morphism relating a nonequilibrium steady state to a local equilibrium state is
invertible, the natural non-equilibrium steady states can be regarded as MacLennan-Zubarev
nonequilibrium ensembles. 
At first sight, invertibility of the M\o ller morphisms seems to be too strong since one can easily 
find a counter example. However, for several systems, the division of the whole system without
a finite part (i.e. the case where ${\cal F}_S=\emptyset$) provides invertible M\o ller morphisms
and, thus, we believe that Proposition 2 holds generically provided that the system is 
divided appropriately. 
Also it should be emphasized that Proposition 2 does not exclude the possibility that the
class of natural nonequilibrium steady states is wider than that of MacLennan-Zubarev ensembles
since the generator of the automorphisims defining the steady states takes MacLennan-Zubarev
form only in a subset of its domain.

In the second half, we have shown that the small system coupled with a single reservoir would
follow `thermodynamic' processes and satisfies the Clausius inequality/equality.
However, it should not be regarded as a dynamical proof of the second law of thermodynamics
since we start from canonical ensembles which are very outcome of the second law. 
We think that the importance of this observation lies in the facts that (i) the dynamical
evolution is consistent with thermodynamics, (ii) one can define thermodynamic heat microscopically,
(iii) entropy generation is identified for step-wise evolution, and (iv) a characterization of a
quasistatic change is given.

\section*{Acknowledgments}
The author thanks Professors L. Accardi, P. Gaspard, H. Hayakawa, C. Jarzynski, 
M.~Ohya, I.~Ojima, 
S. Sasa, K. Sekimoto, A. Shimizu, S. Takesue, Hal Tasaki for discussions
and comments.
This work is partially supported by Grant-in-Aid for Scientific 
Research (C) from the Japan Society of the Promotion of Science, 
by a Grant-in-Aid for Scientific Research of Priority Areas 
``Control of Molecules in Intense Laser Fields'' 
and the 21st Century COE Program at Waseda University ``Holistic Research 
and Education Center for Physics of Self-organization Systems''
both from the Ministry of Education, Culture, Sports, Science and 
Technology of Japan.

}
{

}
}

\end{document}